\documentclass[journal]{IEEEtran}

\usepackage{graphicx}

\usepackage{times}

\usepackage{mathtools}
\usepackage{amsmath}
\usepackage{amssymb}

\usepackage{bbm}

\usepackage{algorithm}
\usepackage{algorithmicx}
\usepackage{algpseudocode}

\usepackage{amsthm}

\usepackage{tikz}
\usetikzlibrary{positioning,calc}

\DeclareMathOperator*{\argmax}{arg\,max}
\DeclareMathOperator*{\argmin}{arg\,min}

\algrenewcommand\algorithmicrequire{\textbf{Input:}}
\algrenewcommand\algorithmicensure{\textbf{Output:}}

\newtheorem{example}{Example}

\usepackage{url}
\usepackage{hyperref}
\usepackage{float}
\usepackage{xcolor}

\usepackage{geometry}
\date{}
\begin{document}

\title{Learning-Based List Sequential Belief Propagation Decoding of Quantum LDPC Codes}


\author{Mohsen~Moradi\textsuperscript{\href{https://orcid.org/0000-0001-7026-0682}{\includegraphics[scale=0.06]{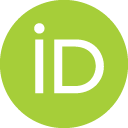}}},
Taejoon~Kim\textsuperscript{\href{https://orcid.org/0000-0002-4017-9530}{\includegraphics[scale=0.06]{figs/ORCID}}},
and 
R\'emi~A.~Chou\textsuperscript{\href{https://orcid.org/0000-0003-4431-3175}{\includegraphics[scale=0.06]{figs/ORCID}}}\\
\thanks{Mohsen Moradi and Taejoon~Kim are with the School of Electrical, Computer and Energy Engineering, Arizona State University, Tempe, AZ 85287, USA (e-mail: mmorad11@asu.edu, taejoonkim@asu.edu).}
\thanks{R\'emi~A.~Chou is with the Department of Computer Science and Engineering, 
The University of Texas at Arlington, Arlington, TX 76019, USA (e-mail: remi.chou@uta.edu).}
\thanks{This work is in part supported by the National Science Foundation (NSF) under grants CNS2451268, CNS2514415, and ITE2515378, and the Office of Naval Research (ONR) under Grant N000142112472.}
}

\maketitle

\begin{abstract}
Quantum low-density parity-check (QLDPC) codes are strong candidates for fault-tolerant quantum computation, but efficient decoding remains a major challenge due to short cycles, degeneracy, and the poor convergence of standard belief-propagation (BP) decoders. We propose a reinforcement learning-based list sequential (RL-LS) BP decoder for QLDPC codes by extending the reinforcement-learning-based sequential variable-node scheduling (RL-S) framework with list-based search. 
At each step, the learned policy selects the next variable node to update; the decoder then retains the ordinary RL-S trajectory while also exploring a competing branch obtained by softly biasing the post-update LLR pair toward the second-most likely Pauli symbol, recomputing the incident local BP messages, and setting the visited variable node to that second-best symbol.
Candidate trajectories are ranked and pruned using our proposed cumulative path metric. The resulting decoder extends the learned decoder by combining the improved convergence of learned sequential scheduling with list exploration. Numerical results on representative QLDPC benchmark codes over the depolarizing channel show that our proposed method improves the decoding performance of the underlying decoder and compares favorably with existing BP-based decoding methods.
\end{abstract}

\begin{IEEEkeywords}
Quantum error correction, quantum LDPC codes, stabilizer codes, CSS codes, belief propagation, quaternary BP, depolarizing channel, sequential scheduling, reinforcement learning, list decoding.
\end{IEEEkeywords}

\section{Introduction}

Quantum error correction is essential for quantum computation, since even small physical error rates can accumulate and corrupt quantum information without active protection \cite{shor1995scheme,terhal2015quantum}. Among the many code families proposed for fault tolerance, quantum low-density parity-check (QLDPC) codes are particularly attractive because their sparse stabilizer structure supports low-weight syndrome extraction while allowing the encoding of many logical qubits \cite{breuckmann2021qldpc}. Over the last several years, progress in QLDPC code design has been substantial. 
Hypergraph-product codes provide explicit positive-rate Calderbank--Shor--Steane (CSS) families whose distance scales proportionally to the square root of the block length \cite{tillich2014quantum}, while later constructions such as balanced-product codes \cite{breuckmann2021balanced}, lifted-product/asymptotically good QLDPC codes \cite{panteleev2022asymptotically}, and quantum Tanner codes \cite{leverrier2022quantumtanner} have significantly improved the asymptotic rate-distance tradeoff and broadened the design space of practical QLDPC families. 
In addition, recent bivariate bicycle (BB) codes have provided concrete finite-length QLDPC examples with strong performance and favorable overhead \cite{bravyi2024high}. 
Recent high-rate QLDPC constructions based on controlled permutation-matrix structures have further expanded the design landscape \cite{kasai2026breaking}. Hardware-aware implementations using reconfigurable atom arrays also highlight the potential of such codes for low-overhead fault-tolerant architectures \cite{zhao2026towards}.
As a result, QLDPC codes are now widely viewed as a leading route toward low-overhead fault-tolerant quantum architectures \cite{breuckmann2021qldpc,roffe2020landscape,bravyi2024high,kasai2026breaking,zhao2026towards}.

Belief propagation (BP) is a natural decoding candidate because of its low complexity, local message-passing structure, and successful use in classical LDPC decoding \cite{gallager1962ldpc,mackay1996near,kschischang2001factor,richardson2001capacity}. For QLDPC codes, however, standard BP is well known to perform poorly \cite{poulin2008iterative,roffe2020landscape}. One reason is structural: the commutativity constraints of stabilizer codes induce many short cycles in the Tanner graph, violating the independence assumptions underlying BP. A second difficulty is uniquely quantum: \emph{degeneracy}, whereby distinct Pauli error patterns can yield the same syndrome and even represent equivalent recovery operations up to stabilizers. Consequently, successful decoding does not require identifying the exact physical error, but rather producing an estimate in the correct logical coset \cite{poulin2008iterative,kuo2022degeneracy}. These features create symmetric pseudo-codewords, trapping sets, and unstable message-passing dynamics, so that plain BP often oscillates, fails to converge, or converges to a syndrome-consistent but logically incorrect solution \cite{poulin2008iterative,kuo2022degeneracy,raveendran2021trapping}.

A large body of work has therefore focused on improving BP-based decoding for QLDPC codes. One direction augments BP with a stronger post-processing stage. Belief propagation with order-0 ordered statistics decoding (BP-OSD-0) uses BP to generate soft reliabilities and then performs an OSD-based post-processing step to enforce syndrome consistency \cite{roffe2020landscape}. Stabilizer inactivation (BP-SI) reduces the cost of this approach by inactivating a carefully selected subset of qubits and solving only a much smaller linear system \cite{ducrest2022si}. More recently, localized statistics decoding (BP-LSD) has shown that much of the post-processing can be confined to small local subgraphs, thereby improving parallelizability while retaining strong decoding performance \cite{hillmann2025lsd}.

A second line of work seeks to improve BP by modifying the message-passing dynamics themselves rather than appending a heavy post-processing stage. Early examples include random perturbation methods for sparse quantum codes \cite{poulin2008iterative}. More recent examples include overcomplete-check and neural BP decoders \cite{miao2023neural,miao2025quaternary}, graph-neural enhancements to iterative decoding \cite{gong2024gnn}, trapping-set-aware message-passing refinements \cite{chytas2024trapping}, and guided-decimation approaches that interleave short BP runs with reliability-based variable fixing \cite{yao2024bpgd}. Related scheduling-oriented advances include layered/random-order decoding and informed dynamic scheduling, both of which exploit the fact that the order in which messages are updated can significantly affect convergence on loopy QLDPC Tanner graphs \cite{ducrest2023layered,moradi2026seqbp, huang2026ids}. 
In parallel, alternative candidate-generation and search-based BP extensions have also begun to emerge, including speculative syndrome-flip post-processing \cite{wang2025bpsf}, multiple-bases BP list decoding \cite{rabeti2025mbbpld}, and learned parallel bit-flipping continuations of learned sequential BP \cite{moradi2026rlsbf}.
Taken together, these results show that breaking the symmetry of plain BP---whether through post-processing, controlled perturbation, decimation, bit flipping, or improved scheduling---is often the key to practical performance gains.

Among these ideas, changing the BP update schedule is especially attractive because it preserves the basic and low complexity message-passing framework and can be implemented with modest overhead. In classical LDPC decoding, shuffled or serial schedules are known to improve convergence \cite{zhang2005shuffled}. Recent work has shown that the same principle is also effective for QLDPC codes. In particular, sequential check-node scheduling (SCNS) and sequential variable-node scheduling (SVNS) can reduce non-convergence and improve error-correction performance relative to conventional flooding BP \cite{moradi2026seqbp}. Building on this idea, reinforcement-learning-based sequential scheduling (RL-S) framework shows that a learned policy can further improve performance by selecting the next variable-node update based on a local syndrome-driven state representation \cite{moradi2026rlbp}. Under depolarizing noise, the resulting quaternary RL-S decoder achieves substantial performance gains over conventional quaternary BP and competitive performance relative to other BP-based baselines \cite{moradi2026rlbp}. RL-based sequential scheduling has also been studied for classical error-correcting codes, including sparse graph-based codes and polar codes \cite{habib2021reldec,moradi2025polar}.

These developments motivate the central question of this paper: can one preserve the convergence gains of learned sequential scheduling while reducing the risk of an early mistake? This question is particularly relevant in quaternary BP under depolarizing noise. At a visited qubit, the local posterior is often not sharply concentrated on a single Pauli symbol, and due to degeneracy a locally non-favorable choice may still belong to a globally successful decoding trajectory. A sequential decoder has no mechanism to revisit such near-misses. This suggests that a soft, list-based extension of learned sequential BP may offer a more flexible way to navigate the local ambiguities induced by degeneracy.

In this paper, we address this gap by proposing a \emph{learning-based list sequential} (RL-LS) BP decoder for QLDPC codes. Our proposed decoder extends the RL-S framework of \cite{moradi2026rlbp} by maintaining a list of candidate trajectories rather than a single trajectory. At each decoding step, a learned policy selects the next variable node to update. 
The ordinary RL-S update is always retained as one candidate, while an additional candidate is generated by softly biasing the post-update LLR pair toward the second-most likely Pauli symbol, recomputing the corresponding incident local BP messages, and setting the visited variable node to that second-best symbol.
Candidate trajectories are ranked and pruned using our proposed cumulative path metric, yielding a decoder that combines learned sequential scheduling with lightweight list exploration while remaining fully within the BP message-passing framework. Because the ordinary RL-S path is always preserved, our proposed decoder is a strict extension of the underlying learned decoder.

A related approach is learned parallel bit flipping for RL-S (RL-S-BF), which first runs an initial RL-S prefix for \(T_0\) iterations and then tests several forced quaternary Pauli changes through independent parallel continuations \cite{moradi2026rlsbf}. While this strategy introduces trajectory diversity, its initial RL-S prefix contributes directly to the total decoding depth. In contrast, the RL-LS decoder proposed here introduces list expansion within the sequential update process itself by retaining the ordinary RL-S continuation and a soft second-symbol continuation at each selected variable node.

In classical coding, a useful perspective is to view decoding as a search over a tree or trellis representation of the code, where each complete path corresponds to a candidate information sequence and hence to a codeword. This viewpoint underlies several important decoding paradigms. In sequential decoding of convolutional codes, Fano decoding and stack decoding adapt the amount of search to the observed channel output and use a path metric to rank partial paths and guide the search toward the transmitted sequence \cite{fano1963heuristic,zigangirov1966some,jelinek1969fast}. In list decoding, multiple candidate paths are retained concurrently, and the search breadth is controlled by the list size; this principle has been especially influential for polar codes \cite{tal2015list}. Similar metric-guided sequential search ideas have also proved effective for polarization-adjusted convolutional (PAC) codes \cite{moradi2021metric}. 
Our proposed RL-LS decoder can also be interpreted as exploring candidate decoding paths, following the same general principle of maintaining promising partial paths and using a path metric to determine which ones should be explored further.

The remainder of the paper is organized as follows. Section~\ref{sec:Background} reviews the necessary background on stabilizer and CSS QLDPC codes, syndrome decoding, degeneracy, quaternary BP under depolarizing noise, and learned sequential scheduling. Section~\ref{sec:RLLS} presents our proposed RL-LS decoder and its path metric. Section~\ref{sec:Implementation} discusses implementation aspects and complexity. Section~\ref{sec:Numerical} provides numerical results on representative QLDPC benchmark codes. Section~\ref{sec:conclusion} concludes the paper.

\section{Background}
\label{sec:Background}

\subsection{Stabilizer and CSS QLDPC codes}
\label{subsec:background_stabilizer}

A single qubit is described by a unit vector in the two-dimensional Hilbert
space \(\mathbb C^2\). The four Pauli operators acting on one qubit are
\[
\begin{aligned}
I&=
\begin{bmatrix}
1 & 0\\
0 & 1
\end{bmatrix},
\qquad
X=
\begin{bmatrix}
0 & 1\\
1 & 0
\end{bmatrix},\\[1ex]
Y&=
\begin{bmatrix}
0 & -i\\
i & 0
\end{bmatrix},
\qquad
Z=
\begin{bmatrix}
1 & 0\\
0 & -1
\end{bmatrix}.
\end{aligned}
\]
Physically, \(X\) corresponds to a bit-flip error, \(Z\) to a phase-flip error,
and \(Y=iXZ\) to a simultaneous bit- and phase-flip error. An \(n\)-qubit system
has state space \((\mathbb C^2)^{\otimes n}\), and an operator of the form
\(P_1\otimes P_2\otimes\cdots\otimes P_n\) means that the single-qubit operator
\(P_i\) acts on qubit \(i\). Let \(\omega\in\{\pm1,\pm i\}\) denote a global
phase. The \(n\)-qubit Pauli group is
\[
\mathcal P_n
=
\left\{
\omega P_1\otimes P_2\otimes\cdots\otimes P_n:
P_i\in\{I,X,Y,Z\}
\right\}.
\]
Since global phases do not affect physical measurement outcomes, Pauli errors
are typically identified with their tensor-product part alone
\cite{gottesman1997stabilizer,calderbank1996good,steane1996multiple}.

An \([[n,k,d]]\) quantum stabilizer code encodes \(k\) logical qubits into
\(n\) physical qubits by specifying an abelian subgroup
\(\mathcal S\subseteq\mathcal P_n\) that does not contain \(-I^{\otimes n}\),
where \(I^{\otimes n}=I\otimes I\otimes\cdots\otimes I\) denotes the identity
operator on the \(n\)-qubit Hilbert space. The subgroup \(\mathcal S\) is called
the stabilizer group. Its elements are called stabilizers or stabilizer
operators, and a set of independent elements whose products generate all of
\(\mathcal S\) is called a set of stabilizer generators. The code space is the
joint \(+1\) eigenspace of all stabilizers,
\[
\mathcal C
=
\left\{
|\psi\rangle\in(\mathbb C^2)^{\otimes n}:
M|\psi\rangle=|\psi\rangle,\ \forall M\in\mathcal S
\right\}.
\]
Equivalently, \(\mathcal C\) is the set of all \(n\)-qubit states that satisfy all
stabilizer constraints. Since every stabilizer can be written as a product of
stabilizer generators, it is enough in practice to specify and measure only the
generators. The minimum distance \(d\) is the minimum weight of a Pauli operator
in \(N(\mathcal S)\setminus\mathcal S\), where
\[
N(\mathcal S)
=
\{P\in\mathcal P_n:\; PS=SP \text{ for all } S\in\mathcal S\}
\]
is the normalizer of \(\mathcal S\). Operators in
\(N(\mathcal S)\setminus\mathcal S\) preserve the code space but act
nontrivially on the encoded logical qubits \cite{gottesman1997stabilizer}.

In the binary symplectic formalism, an \(m\)-generator stabilizer code is
represented by
\[
H=[\,H_X\;H_Z\,]\in\mathbb F_2^{m\times 2n},
\]
where each row describes the \(X\)-support and \(Z\)-support of one stabilizer
generator. The commutativity of the generators is equivalent to the symplectic
orthogonality condition
\[
H_XH_Z^\top+H_ZH_X^\top=0
\qquad \text{over } \mathbb F_2.
\]
This binary representation is especially useful for decoding, because syndrome
formation and message-passing updates can then be expressed using sparse binary
matrices \cite{gottesman1997stabilizer,breuckmann2021qldpc}.

In this paper, we focus on CSS QLDPC codes. A CSS code is specified by two
sparse binary matrices
\[
H_X\in\mathbb F_2^{m_X\times n},
\qquad
H_Z\in\mathbb F_2^{m_Z\times n},
\]
satisfying
\[
H_XH_Z^\top=0.
\]
The rows of \(H_X\) define the \(X\)-type stabilizer checks, and the rows of
\(H_Z\) define the \(Z\)-type stabilizer checks. Because both matrices are
sparse, each qubit participates in only a small number of checks and each check
acts on only a small number of qubits. This low-density structure is what makes
QLDPC codes attractive for scalable syndrome extraction and iterative
message-passing decoding \cite{calderbank1996good,steane1996multiple,breuckmann2021qldpc}.

A convenient graphical representation of a CSS code can be given by two Tanner
graphs, one associated with \(H_X\) and one associated with \(H_Z\). In each
graph, the variable nodes correspond to physical qubits and the check nodes
correspond to stabilizer checks. An edge is present whenever a qubit participates
in a check. These sparse bipartite graphs provide the computational structure on
which the BP-based decoders in this paper operate.

\subsection{Syndrome decoding, logical equivalence, and degeneracy}
\label{subsec:background_degeneracy}

Let
\[
Q=(Q_1,\ldots,Q_n),\qquad Q_i\in\{I,X,Y,Z\},
\]
denote an \(n\)-qubit Pauli error acting on a CSS code. It is convenient to
decompose \(Q\) into two binary component vectors. For each single-qubit Pauli
symbol \(q\in\{I,X,Y,Z\}\), define
\[
e^X(q)\triangleq \mathbbm{1}[q\in\{X,Y\}],
\qquad
e^Z(q)\triangleq \mathbbm{1}[q\in\{Y,Z\}].
\]
Thus, \(e^X(q)\) is the binary \(X\)-component of the error and is detected by
\(H_Z\), whereas \(e^Z(q)\) is the binary \(Z\)-component of the error and is
detected by \(H_X\). For the full Pauli error \(Q\), these definitions induce
\[
\begin{aligned}
\mathbf e^X&=(e^X(Q_1),\ldots,e^X(Q_n)),\\
\mathbf e^Z&=(e^Z(Q_1),\ldots,e^Z(Q_n)).
\end{aligned}
\]
The measured CSS syndrome is
\[
\mathbf s^X=H_X\mathbf e^Z,
\qquad
\mathbf s^Z=H_Z\mathbf e^X,
\]
where all operations are over \(\mathbb F_2\). In other words, \(X\)-type checks
are triggered by the \(Z\)-component of the error, while \(Z\)-type checks are
triggered by the \(X\)-component
\cite{gottesman1997stabilizer,calderbank1996good,steane1996multiple}.

Given the measured syndrome \((\mathbf s^X,\mathbf s^Z)\), a decoder outputs a
Pauli estimate
\[
\hat Q=(\hat Q_1,\ldots,\hat Q_n),
\]
or equivalently the induced binary component estimates
\((\hat{\mathbf e}^X,\hat{\mathbf e}^Z)\). For a tentative estimate, define the
residual mismatch vectors
\[
\boldsymbol\delta^X
\triangleq
\mathbf s^X\oplus H_X\hat{\mathbf e}^Z,
\qquad
\boldsymbol\delta^Z
\triangleq
\mathbf s^Z\oplus H_Z\hat{\mathbf e}^X,
\]
and the residual mismatch weight
\[
w\triangleq \|\boldsymbol\delta^X\|_1+
\|\boldsymbol\delta^Z\|_1.
\]
The condition \(w=0\) means that the tentative estimate satisfies the measured
syndrome. However, syndrome satisfaction alone is not sufficient for successful
quantum decoding. For a classical linear code, the natural objective is to find
the most likely error vector consistent with the syndrome. In the quantum
setting, the objective is subtler: because of degeneracy, different physical
Pauli errors can have the same syndrome and can also be logically equivalent if
they differ by a stabilizer.

More precisely, if the actual channel error is \(Q\) and the decoder outputs
\(\hat Q\), then there are three relevant cases:
\begin{itemize}
    \item if \(\hat Q Q\in\mathcal S\), then decoding succeeds, even if
    \(\hat Q\neq Q\);
    \item if \(\hat Q Q\in N(\mathcal S)\setminus\mathcal S\), then a logical
    error has occurred;
    \item if \(\hat Q\) is not syndrome-consistent, then decoding has failed to
    return a valid recovery.
\end{itemize}
Thus, successful decoding means returning a recovery in the correct logical
coset, not necessarily recovering the exact physical error
\cite{poulin2008iterative,kuo2022degeneracy,gottesman1997stabilizer}.

Degeneracy is one of the main reasons why plain BP is less reliable for QLDPC
codes than for classical LDPC codes. It creates multiple competing explanations
of the same syndrome, making the posterior landscape flatter and more symmetric.
At the same time, QLDPC Tanner graphs contain many short cycles, so these
competing explanations can reinforce one another through loopy message passing.
As a result, BP may oscillate, fail to converge, or converge to a
syndrome-consistent but logically incorrect estimate
\cite{poulin2008iterative,kuo2022degeneracy,raveendran2021trapping}. Throughout
this paper, decoding performance is measured by the \emph{block error rate},
i.e., the probability that the decoder either fails to return a syndrome-consistent
estimate or returns an estimate that is not logically equivalent to the actual
channel error.

\subsection{Belief propagation for CSS codes under depolarizing noise}
\label{subsec:background_qbp}

Belief propagation is an iterative message-passing algorithm on the Tanner graph
\cite{kschischang2001factor}. In the CSS setting, BP operates on two Tanner
graphs, one associated with \(H_X\) and one associated with \(H_Z\). Since
\(X\)-type checks detect the \(Z\)-component of the error and \(Z\)-type checks
detect the \(X\)-component, the \(H_X\) graph is used to infer
\(\mathbf e^Z\), while the \(H_Z\) graph is used to infer \(\mathbf e^X\). For
qubit \(v_i\), define
\[
\begin{aligned}
\mathcal N_X(v_i)&\triangleq \{a:\;H_X(a,i)=1\},\\
\mathcal N_Z(v_i)&\triangleq \{b:\;H_Z(b,i)=1\},
\end{aligned}
\]
where \(\mathcal N_X(v_i)\) and \(\mathcal N_Z(v_i)\) are the neighboring checks of
qubit \(v_i\) in the two Tanner graphs. In syndrome-domain BP, each edge carries
variable-to-check and check-to-variable messages. The check-node updates retain
the usual sum-product form because each syndrome constraint is binary
\cite{poulin2008iterative,mackay1999good,kschischang2001factor}.

Under an independent Pauli-\(X\) or Pauli-\(Z\) channel, the corresponding binary
component can be decoded using standard binary BP. In this paper, however, the
main focus is the depolarizing channel, where \(p\) denotes the physical error
probability and each qubit satisfies
\[
\begin{aligned}
\Pr(Q_i=I)&=1-p,\\
\Pr(Q_i=X)&=\Pr(Q_i=Y)=\Pr(Q_i=Z)=\frac{p}{3}.
\end{aligned}
\]
Under this model, the \(X\)- and \(Z\)-components are statistically coupled
through the possibility of a \(Y\) error. Therefore, decoding the two components
independently is suboptimal, and it is more natural to work with a quaternary
posterior over \(\{I,X,Y,Z\}\)
\cite{poulin2008iterative,miao2025quaternary,moradi2026rlbp}.

The quaternary-coupled implementation used in this paper keeps two scalar
log-likelihood ratio (LLR) streams. Let \(L_i^X\) denote the a-posteriori LLR
for the \(X\)-component of qubit \(v_i\), obtained from the \(H_Z\) graph, and let
\(L_i^Z\) denote the a-posteriori LLR for the \(Z\)-component, obtained from the
\(H_X\) graph. These two LLRs are then combined to reconstruct the four-symbol
belief at the variable node. For the depolarizing prior, define
\[
\pi_I=1-p,
\qquad
\pi_X=\pi_Y=\pi_Z=\frac p3.
\]
For \(q\in\{I,X,Y,Z\}\), define the depolarizing correction factor
\[
\kappa_q(p)\triangleq
\begin{cases}
\dfrac{\pi_I}{\left(1-\frac{2p}{3}\right)^2}, & q=I,\\[1.4ex]
\dfrac{\pi_X}{\left(1-\frac{2p}{3}\right)\frac{2p}{3}}, & q=X,\\[1.4ex]
\dfrac{\pi_Y}{\left(\frac{2p}{3}\right)^2}, & q=Y,\\[1.4ex]
\dfrac{\pi_Z}{\left(1-\frac{2p}{3}\right)\frac{2p}{3}}, & q=Z.
\end{cases}
\]
The unnormalized quaternary belief at node \(i\) can be written explicitly as \cite{moradi2026rlbp}
\[
\tilde B_i(q)
\triangleq
\begin{cases}
\kappa_I(p)\exp\!\left(\dfrac{L_i^Z+L_i^X}{2}\right), & q=I,\\[1.4ex]
\kappa_X(p)\exp\!\left(\dfrac{L_i^Z-L_i^X}{2}\right), & q=X,\\[1.4ex]
\kappa_Y(p)\exp\!\left(\dfrac{-L_i^Z-L_i^X}{2}\right), & q=Y,\\[1.4ex]
\kappa_Z(p)\exp\!\left(\dfrac{-L_i^Z+L_i^X}{2}\right), & q=Z.
\end{cases}
\]
The normalized quaternary belief is
\[
B_i(q)
=
\frac{\tilde B_i(q)}
{\sum_{q'\in\{I,X,Y,Z\}}\tilde B_i(q')},
\]
and the ordinary local hard decision is
\[
\hat q_i
=
\argmax_{q\in\{I,X,Y,Z\}}B_i(q).
\]
The explicit four-symbol expression above is useful for seeing how the two LLR
streams combine: a positive \(L_i^X\) favors no \(X\)-component, a negative
\(L_i^X\) favors an \(X\)-component, and similarly for \(L_i^Z\). The correction
factor \(\kappa_q(p)\) then restores the depolarizing-channel coupling between
the four Pauli symbols. This two-stream representation preserves the
low-complexity message-passing structure of BP while accounting for the coupling
induced by depolarizing noise \cite{miao2025quaternary,moradi2026rlbp}.

Although BP is attractive because of its locality and low complexity, its
performance on QLDPC codes is often degraded by short cycles, degeneracy, and
unstable iterative dynamics. These issues motivate schedule-based and list-based
modifications that retain the BP message-passing structure while improving
convergence and reliability.

\subsection{Sequential scheduling and learned update order}
\label{subsec:background_scheduling}

In conventional flooding BP, all check-to-variable messages are updated in
parallel using the previous iteration's variable-to-check messages, and then all
variable-to-check messages are updated in parallel. This schedule is simple and
highly parallelizable, but on Tanner graphs with many short cycles it can
reinforce oscillatory or ambiguous beliefs and slow the propagation of reliable
local information \cite{zhang2005shuffled,moradi2026seqbp}.

Sequential schedules modify this behavior by updating nodes one at a time, so
that newly computed messages are immediately available to later updates in the
same sweep. In sequential check-node scheduling (SCNS), the decoder processes
check nodes in a chosen order. In sequential variable-node scheduling (SVNS), it
processes variable nodes in sequence. Both approaches inject asymmetry into the
message-passing trajectory and can improve convergence by allowing later updates
to exploit fresher information \cite{moradi2026seqbp}. Recent work has shown that such schedules are
effective for QLDPC decoding, where they can reduce non-convergence and improve
error-correction performance relative to conventional flooding BP
\cite{moradi2026seqbp,ducrest2023layered,huang2026ids}.

The choice of which node to update next is also important. A fixed sequential
order can already help, but a state-dependent order can adapt the update order
to the current residual syndrome pattern. This is the key idea behind the
reinforcement-learning-based sequential scheduling (RL-S) framework
\cite{moradi2026rlbp}. In RL-S, the next variable node is selected by a learned
policy that observes a compact local state derived from the residual mismatch
around each candidate node. The policy is trained offline and then frozen at
inference time. Importantly, the underlying BP update equations are unchanged;
the learned component affects only the update order. This modularity makes RL-S
a natural baseline on which to build list-based extensions.

For each variable node \(i\), RL-S forms a compact local state from the residual
mismatch pattern around that node. Assume fixed deterministic edge orderings for
\(\mathcal N_X(v_i)\) and \(\mathcal N_Z(v_i)\). Let \(\beta_X(a,i)\) and
\(\beta_Z(b,i)\) denote the corresponding power-of-two edge weights. Define
\[
\sigma_i^X
\triangleq
\sum_{a\in\mathcal N_X(i)}\boldsymbol\delta^X_a\,\beta_X(a,i),
\qquad
\sigma_i^Z
\triangleq
\sum_{b\in\mathcal N_Z(i)}\boldsymbol\delta^Z_b\,\beta_Z(b,i),
\]
and let
\[
A_{\max}
\triangleq
\max_i\{\,|\mathcal N_X(v_i)|,|\mathcal N_Z(v_i)|\,\}.
\]
The combined local RL state is
\[
\sigma_i
\triangleq
\sigma_i^X+2^{A_{\max}}\sigma_i^Z.
\]
Given the remaining set \(R\) of variable nodes not yet visited in the current
sweep, RL-S selects
\[
u^\star\in\argmax_{u\in R}Q(\sigma_u,u),
\]
where \(Q(\sigma,u)\) is the frozen Q-table learned offline. After updating
\(u^\star\), only the residual checks incident on \(u^\star\) and the local
states of variable nodes in the corresponding second-order neighborhood need to
be refreshed. Thus, RL-S preserves the low-complexity message-passing structure
of BP while using a learned policy to guide the sequential trajectory. Our
proposed RL-LS decoder in the next section keeps this learned scheduler but
replaces the single trajectory with a list of candidate trajectories.

\section{Proposed RL-LS Decoder}
\label{sec:RLLS}

Our proposed RL-LS decoder extends RL-S by keeping a list of candidate decoding
trajectories instead of a single trajectory. Each list element stores the current
quaternary BP messages, the tentative hard decision \(\hat{\mathbf q}\), the
induced binary estimates \((\hat{\mathbf e}^X,\hat{\mathbf e}^Z)\), the residual
mismatch vectors \((\boldsymbol\delta^X,\boldsymbol\delta^Z)\), the mismatch
weight \(w\), the local RL states \(\{\sigma_i\}\), the remaining set of variable
nodes in the current sweep, and a cumulative path metric \(M\). The learned
Q-table is the same as in RL-S; it is used only to select the next variable node
to update inside each surviving path.

The key difference from RL-S is that RL-LS does not always commit to a single
local continuation. At each selected variable node, RL-LS first keeps the
ordinary RL-S continuation. It then creates one additional continuation by
softly biasing the post-update LLR pair toward the second-most likely Pauli
symbol and recomputing the incident local BP messages. The resulting children
are ranked using our proposed cumulative path metric, and only the best \(L\) paths are
retained.

\subsection{Local list expansion}
\label{subsec:local_score}

Let \(\mathcal V_{\mathrm{act}}\) be the set of active variable nodes considered
by the decoder. For a surviving list path \(c_\ell\), let
\(R^{(\ell)}\subseteq\mathcal V_{\mathrm{act}}\) denote the set of nodes not yet
visited in the current sweep. The next node for this path is selected using the
same greedy learned scheduler as RL-S:
\[
u^\star\in\argmax_{u\in R^{(\ell)}} Q(\sigma_u^{(\ell)},u).
\]
The selected node is then removed from \(R^{(\ell)}\). The decoder first applies
the ordinary RL-S local update at \(u^\star\). This produces the ordinary child
path \(c_{\ell,1}\), which contains the updated BP messages, hard decision,
residual mismatches, local RL states, and path metric after the standard RL-S
update. Let \((L_{u^\star}^X,L_{u^\star}^Z)\) denote the post-update
a-posteriori LLR pair at the selected node. From this pair, define the local
log-score
\[
\begin{aligned}
&s_{u^\star}(q)
\triangleq \log \tilde B_{u^\star}(q)\\
&=
\log \kappa_q(p)
+
\frac12\bigl(1-2e^Z(q)\bigr)L_{u^\star}^Z
+
\frac12\bigl(1-2e^X(q)\bigr)L_{u^\star}^X,
\end{aligned}
\]
for \(q\in\{I,X,Y,Z\}\). Let
\[
q_{u^\star,1}
\triangleq
\argmax_{q\in\{I,X,Y,Z\}} s_{u^\star}(q)
\]
and
\[
q_{u^\star,2}
\triangleq
\argmax_{q\in\{I,X,Y,Z\}\setminus\{q_{u^\star,1}\}}
s_{u^\star}(q)
\]
denote the best and second-best Pauli symbols at the selected node. The local
score gap is
\[
\Delta_{u^\star}
\triangleq
s_{u^\star}(q_{u^\star,1})-s_{u^\star}(q_{u^\star,2})\ge 0.
\]
This gap measures how strongly the local quaternary posterior favors the
ordinary hard decision over the closest competing Pauli symbol.

The ordinary child follows the standard RL-S local update and receives no
additional metric penalty:
\[
M^{(\ell,1)}\gets M^{(\ell)}.
\]
To generate the alternative child, define
\[
\alpha_{u^\star}\triangleq \Delta_{u^\star}+\tau,
\]
where \(\tau\geq 0\) is a small fixed margin used to strengthen the soft second-symbol branch (in our simulations, we use \(\tau=0.25\)). The post-update LLR pair is then biased toward the second-best symbol by
\[
\begin{aligned}
\bar L_{u^\star}^X
&\triangleq
L_{u^\star}^X
+
\alpha_{u^\star}\bigl(1-2e^X(q_{u^\star,2})\bigr),\\
\bar L_{u^\star}^Z
&\triangleq
L_{u^\star}^Z
+
\alpha_{u^\star}\bigl(1-2e^Z(q_{u^\star,2})\bigr).
\end{aligned}
\]
Starting from the ordinary child \(c_{\ell,1}\), the decoder recomputes the local
BP messages incident on \(u^\star\) using
\((\bar L_{u^\star}^X,\bar L_{u^\star}^Z)\). It then sets the visited variable
node to \(q_{u^\star,2}\) and updates the induced binary components, the
residual mismatch vectors, the mismatch weight, and the affected local RL
states. The alternative child receives the local metric penalty
\[
M^{(\ell,2)}\gets M^{(\ell)}+\Delta_{u^\star}.
\]
The motivation for this additive metric is its local likelihood-ratio interpretation. Since \(s_u(q)=\log \tilde B_u(q)\), the score gap can be written
as
\[
\Delta_u
=
\log
\frac{\tilde B_u(q_{u,1})}{\tilde B_u(q_{u,2})}
=
\log
\frac{B_u(q_{u,1})}{B_u(q_{u,2})},
\]
where the normalization cancels in the second equality. Hence, \(\Delta_u\) is the local log-likelihood penalty for replacing the locally dominant Pauli symbol by the closest competing symbol. If \(\mathcal D^{(\ell)}\) denotes the set of local expansion steps at which path \(c_\ell\) takes the second-symbol branch, then
\[
\begin{aligned}
M^{(\ell)}
=\sum_{r\in\mathcal D^{(\ell)}} \Delta_{u_r} 
=
- \sum_{r\in\mathcal D^{(\ell)}}\log
\frac{B_{u_r}(q_{u_r,2})}{B_{u_r}(q_{u_r,1})},
\end{aligned}
\]
where \(u_r\) is the variable node selected at step \(r\). Therefore, smaller values of \(M\) correspond to paths that deviate from the ordinary RL-S trajectory only at locally ambiguous nodes, while larger values correspond to deviations against stronger local BP evidence.


\subsection{List pruning and final decision}
\label{subsec:list_pruning}

After the current list elements have been expanded, the decoder collects the
resulting children into a temporary set \(\mathcal C\). Since the list cannot be
allowed to grow without bound, RL-LS retains only the best
\[
\min(L,|\mathcal C|)
\]
children, where \(L\) is the maximum list size. Candidates with smaller path
metric \(M\) are preferred; ties are broken by the smaller residual mismatch
weight \(w\). Thus, pruning first favors paths that remain close to the locally
dominant quaternary choices and then, among paths with equal metric, favors
those satisfying more residual checks.

If any list element reaches \(w=0\), the decoder terminates and returns that
candidate as a syndrome-consistent decoding outcome. If no path reaches
\(w=0\) within the allowed number of iterations, the decoder returns the
surviving path with the smallest \((M,w)\) pair in lexicographic order and
declares non-convergence. Algorithm~\ref{alg:rlls} summarizes our proposed
RL-LS decoder.

\begin{algorithm}[!t]
\small
\caption{RL-LS decoding with soft second-symbol expansion}
\label{alg:rlls}
\begin{algorithmic}[1]
\Require Q-table \(Q(\sigma,u)\), maximum list size \(L\), maximum number of iterations \(I_{\max}\), perturbation parameter \(\tau\)
\Ensure Estimated Pauli error \(\hat{\mathbf q}\), induced binary estimates \((\hat{\mathbf e}^X,\hat{\mathbf e}^Z)\), and convergence flag

\State Initialize the root list element \(c_1\) with the ordinary RL-S state:
BP messages, hard decision, residual mismatches, local RL states,
mismatch weight \(w^{(1)}\), remaining set, and metric \(M^{(1)}\gets0\)
\State Set \(\mathcal L\gets\{c_1\}\)

\For{\(t=1\) to \(I_{\max}\)}
    \ForAll{\(c_\ell\in\mathcal L\)}
        \State Reset the remaining set \(R^{(\ell)}\gets\mathcal V_{\mathrm{act}}\)
    \EndFor

    \While{there exists \(c_\ell\in\mathcal L\) with \(R^{(\ell)}\neq\emptyset\)}
        \State Initialize \(\mathcal C\gets\emptyset\)

        \ForAll{\(c_\ell\in\mathcal L\)}
            \If{\(w^{(\ell)}=0\)}
                \State \Return the estimate of \(c_\ell\) with convergence flag \(1\)
            \EndIf

            \If{\(R^{(\ell)}=\emptyset\)}
                \State Add \(c_\ell\) to \(\mathcal C\) and continue
            \EndIf

            \State Select the next variable node
            \[
            u^\star\in\argmax_{u\in R^{(\ell)}} Q(\sigma_u^{(\ell)},u)
            \]
            \State Remove \(u^\star\) from \(R^{(\ell)}\)

            \State \textbf{Ordinary child:} copy \(c_\ell\) into \(c_{\ell,1}\)
            \State Apply one ordinary RL-S local update at \(u^\star\), obtaining \(c_{\ell,1}\) and \((L_{u^\star}^X,L_{u^\star}^Z)\)
            \State Compute \(s_{u^\star}(q)\) for \(q\in\{I,X,Y,Z\}\)
            \State Find \(q_{u^\star,1}\), \(q_{u^\star,2}\), and \(\Delta_{u^\star}\)
            \State Set \(M^{(\ell,1)}\gets M^{(\ell)}\)
            \State Add \(c_{\ell,1}\) to \(\mathcal C\)

            \State \textbf{Alternative child:} copy the updated state \(c_{\ell,1}\) into \(c_{\ell,2}\)
            \State Set \(\alpha_{u^\star}\gets\Delta_{u^\star}+\tau\)
            \State Perturb the updated pair toward \(q_{u^\star,2}\):
            \[
            \begin{aligned}
            \bar L_{u^\star}^X
            &\gets L_{u^\star}^X+\alpha_{u^\star}(1-2e^X(q_{u^\star,2})),\\
            \bar L_{u^\star}^Z
            &\gets L_{u^\star}^Z+\alpha_{u^\star}(1-2e^Z(q_{u^\star,2}))
            \end{aligned}
            \]
            \State Recompute the incident local BP messages of \(u^\star\) using \((\bar L_{u^\star}^X,\bar L_{u^\star}^Z)\)
            \State Set the Pauli symbol at \(u^\star\) to \(q_{u^\star,2}\)
            \State Update the induced binary components, residual mismatches, mismatch weight, and affected RL states
            \State Set \(M^{(\ell,2)}\gets M^{(\ell)}+\Delta_{u^\star}\)
            \State Add \(c_{\ell,2}\) to \(\mathcal C\)
        \EndFor

        \State Keep the best \(\min(L,|\mathcal C|)\) candidates according to \((M,w)\)
        \State Set \(\mathcal L\) equal to these surviving candidates
    \EndWhile
\EndFor

\State Let \(c^\star\in\argmin_{c_\ell\in\mathcal L}(M^{(\ell)},w^{(\ell)})\) in lexicographic order
\State \Return the estimate of \(c^\star\) with convergence flag \(0\)
\end{algorithmic}
\end{algorithm}

Algorithm~\ref{alg:rlls} can be read as a list-based extension of the ordinary
RL-S decoder. The algorithm has four main stages: initialization, sweep-level
processing, local list expansion, and pruning/final selection. The list
\(\mathcal L\) contains the currently surviving decoding trajectories, while
\(\mathcal C\) is a temporary set used to collect the children generated at the
current local expansion step. Each path in the list carries its own BP messages,
hard decision, residual mismatches, local RL states, remaining set of unvisited
variable nodes, residual mismatch weight, and cumulative path metric.

Lines~1--2 initialize the decoder. The root path \(c_1\) is initialized exactly
as in the ordinary RL-S decoder. Its path metric is set to zero because no
second-symbol deviation has been taken yet, and the initial list therefore
contains only this root path.

Lines~3--6 start a new decoding iteration. At the beginning of each iteration,
every surviving path resets its remaining set \(R^{(\ell)}\) to the active
variable-node set. Thus, within the current sweep, each path visits active
variable nodes sequentially according to the learned RL-S scheduler. The
difference from ordinary RL-S is that this reset is performed for every
surviving list path rather than for a single trajectory.

Lines~7--8 begin the local expansion loop. This loop continues as long as at
least one path still has an unvisited variable node in the current sweep. At each
local expansion step, the temporary child set \(\mathcal C\) is cleared, and the
children generated from all current paths are inserted into this set before
pruning.

Lines~9--17 process each current path \(c_\ell\). If a path has already reached
zero residual mismatch weight, then it is syndrome-consistent and the decoder
immediately returns its estimate. If the path has no remaining variable node to
visit in the current sweep, it is copied unchanged into the temporary set.
Otherwise, the learned Q-table is used exactly as in RL-S to select the next
variable node \(u^\star\) from the remaining set of that path. The selected node
is then removed from \(R^{(\ell)}\), so it will not be visited again in the same
sweep for that path.

Lines~18--23 generate the ordinary child \(c_{\ell,1}\). This child follows the
standard RL-S continuation: the decoder copies the current path, applies one
ordinary quaternary SVNS update at the selected variable node, updates the BP
messages and hard decision, and obtains the post-update LLR pair at that node.
The local Pauli scores are then computed from this post-update pair, and the
best and second-best Pauli symbols are identified. Since this child follows the
ordinary RL-S decision, its path metric is unchanged. Thus, the ordinary RL-S
trajectory is always preserved inside the list.

Lines~24--31 generate the second-symbol child \(c_{\ell,2}\). This child starts from the ordinary child \(c_{\ell,1}\), but then softly biases the post-update LLR pair toward the second-best Pauli symbol. The incident BP messages of the selected variable node are recomputed using this perturbed pair, and the hard decision at the selected node is set to the second-best symbol. The metric of this child is increased by the local score gap, so that deviations from a strongly preferred ordinary decision are penalized more than deviations between two nearly tied Pauli symbols.

Lines~33--34 perform list pruning. After all current paths have been expanded, the temporary set \(\mathcal C\) may contain more than \(L\) candidates. The decoder keeps only the best candidates according to the lexicographic pair \((M,w)\). The surviving candidates become the new list \(\mathcal L\), and the next local expansion step continues from this pruned list.

Lines~37--38 describe the fallback output when no path reaches zero residual
mismatch within the allowed number of iterations. In this case, the decoder
selects the surviving path with the smallest lexicographic pair \((M,w)\) and
returns its estimate with a non-convergence flag. Therefore, the returned path is
the one that remains closest to the sequence of locally dominant Pauli decisions
while also having the smallest residual mismatch among paths with the same
metric.

In Example~\ref{ex:toy_setup_rlls}, we first define a small CSS code and compute the corresponding syndrome pair. Then, in Example~\ref{ex:rlls_step}, we use the same instance to trace one local expansion step of Algorithm~\ref{alg:rlls}. Together, the two examples illustrate how our proposed RL-LS decoder forms the ordinary RL-S child, constructs the second-symbol child, and prunes the resulting candidate paths.

\begin{example}
\label{ex:toy_setup_rlls}
Consider the following CSS code with \(n=12\) variable nodes, \(m_X=6\)
\(X\)-type checks, and \(m_Z=6\) \(Z\)-type checks:
\begingroup
\scriptsize
\setlength{\arraycolsep}{3pt}
\[
H_X=
\left[
\begin{array}{*{12}{c}}
1 & 0 & 1 & 0 & 0 & 0 & 1 & 0 & 0 & 1 & 0 & 0 \\
1 & 1 & 0 & 0 & 0 & 0 & 0 & 1 & 0 & 0 & 1 & 0 \\
0 & 1 & 1 & 0 & 0 & 0 & 0 & 0 & 1 & 0 & 0 & 1 \\
0 & 0 & 0 & 1 & 0 & 1 & 1 & 0 & 0 & 1 & 0 & 0 \\
0 & 0 & 0 & 1 & 1 & 0 & 0 & 1 & 0 & 0 & 1 & 0 \\
0 & 0 & 0 & 0 & 1 & 1 & 0 & 0 & 1 & 0 & 0 & 1
\end{array}
\right],
\]
\endgroup
\begingroup
\scriptsize
\setlength{\arraycolsep}{3pt}
\[
H_Z=
\left[
\begin{array}{*{12}{c}}
1 & 0 & 0 & 1 & 0 & 0 & 1 & 1 & 0 & 0 & 0 & 0 \\
0 & 1 & 0 & 0 & 1 & 0 & 0 & 1 & 1 & 0 & 0 & 0 \\
0 & 0 & 1 & 0 & 0 & 1 & 1 & 0 & 1 & 0 & 0 & 0 \\
1 & 0 & 0 & 1 & 0 & 0 & 0 & 0 & 0 & 1 & 1 & 0 \\
0 & 1 & 0 & 0 & 1 & 0 & 0 & 0 & 0 & 0 & 1 & 1 \\
0 & 0 & 1 & 0 & 0 & 1 & 0 & 0 & 0 & 1 & 0 & 1
\end{array}
\right].
\]
\endgroup
These matrices satisfy \(H_XH_Z^\top=0\) over \(\mathbb F_2\). The corresponding
Tanner graph is shown in Fig.~\ref{fig:toy_css_tanner_12_doubleoutline}. In the
figure, the highlighted neighborhoods indicate sample local regions of the graph;
Example~\ref{ex:rlls_step} uses the neighborhood of \(v_7\) to illustrate one
RL-LS expansion step.

\begin{figure*}[t]
\centering
\input{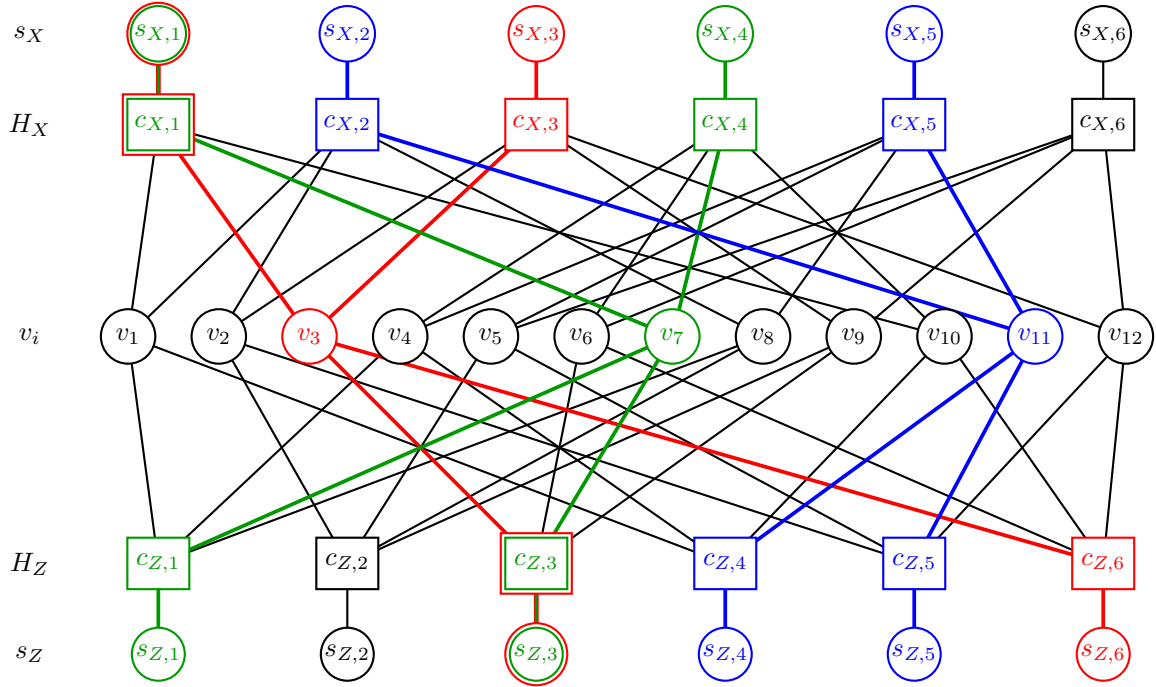}
\caption{Tanner graph for the 12-variable CSS example. The highlighted variable nodes illustrate one RL-LS decoding step: \(v_3\) and \(v_{11}\) have already been explored in the current trajectory, and \(v_7\) is the current variable node where the local list expansion is carried out.}
\label{fig:toy_css_tanner_12_doubleoutline}
\end{figure*}

Suppose that the channel introduces the Pauli error
\[
Q=(I,X,I,Z,Y,I,I,X,Z,I,Y,I).
\]
Using \(e^X(q)=\mathbbm{1}[q\in\{X,Y\}]\) and
\(e^Z(q)=\mathbbm{1}[q\in\{Y,Z\}]\), the induced component vectors are
\[
\mathbf e^X=(0,1,0,0,1,0,0,1,0,0,1,0),
\]
\[
\mathbf e^Z=(0,0,0,1,1,0,0,0,1,0,1,0).
\]
Therefore, the syndrome pair is
\[
\mathbf s^X
=H_X\mathbf e^Z
=
\begin{bmatrix}
0\\1\\1\\1\\1\\0
\end{bmatrix},
\qquad
\mathbf s^Z
=H_Z\mathbf e^X
=
\begin{bmatrix}
1\\1\\0\\1\\1\\0
\end{bmatrix}.
\]
This syndrome pair will be used in Example~\ref{ex:rlls_step} to illustrate one
local RL-LS expansion step.
\end{example}

\begin{example}
\label{ex:rlls_step}
Consider the toy code and syndrome pair from Example~\ref{ex:toy_setup_rlls}.
Focus on one surviving list path \(c_\ell\) during one sweep of
Algorithm~\ref{alg:rlls}. 
Suppose that the local residual pattern around variable node \(v_7\) in this
path is
\[
\bigl[\boldsymbol\delta^X(c_{X,1}),\boldsymbol\delta^X(c_{X,4})\bigr]
=[1,0],
\]
\[
\bigl[\boldsymbol\delta^Z(c_{Z,1}),\boldsymbol\delta^Z(c_{Z,3})\bigr]
=[1,1].
\]
From the Tanner graph,
\[
\mathcal N_X(v_7)=\{c_{X,1},c_{X,4}\},
\qquad
\mathcal N_Z(v_7)=\{c_{Z,1},c_{Z,3}\}.
\]
Since \(A_{\max}=2\), the local binary-to-integer states are
\[
\sigma_7^X=1\cdot 2^0+0\cdot 2^1=1,
\qquad
\sigma_7^Z=1\cdot 2^0+1\cdot 2^1=3.
\]
Thus,
\[
\sigma_7
=
\sigma_7^X+2^{A_{\max}}\sigma_7^Z
=1+2^2\cdot 3
=13.
\]
Assume that the frozen Q-table gives the largest value at this path for
variable node \(7\), i.e.,
\[
7\in\argmax_{u\in R^{(\ell)}}Q(\sigma_u^{(\ell)},u).
\]
Then Algorithm~\ref{alg:rlls} selects \(u^\star=7\), equivalently the variable
node \(v_{u^\star}=v_7\), and removes it from \(R^{(\ell)}\).

The ordinary child \(c_{\ell,1}\) is obtained by applying the usual RL-S local
quaternary SVNS update at \(v_7\). This ordinary update computes the incoming
check-to-variable information from the neighboring checks on both Tanner graphs,
updates the local a-posteriori LLR pair \((L_7^X,L_7^Z)\), forms the local
quaternary belief over \(\{I,X,Y,Z\}\), and makes the corresponding hard
decision. RL-LS uses this ordinary update as the first child and then constructs
a second child from the same post-update local information.

Suppose, for illustration, that this update produces the following post-update
local log-scores:
\[
\begin{array}{c|cccc}
q & I & X & Y & Z \\ \hline
s_7(q) & 1.45 & 0.20 & 1.05 & -0.70
\end{array}
\]
Then
\[
q_{7,1}=I,
\qquad
q_{7,2}=Y,
\qquad
\Delta_7=s_7(I)-s_7(Y)=0.40.
\]
The ordinary child follows the locally dominant Pauli decision and keeps the
same path metric,
\[
M^{(\ell,1)}=M^{(\ell)}.
\]

The alternative child \(c_{\ell,2}\) explores the closest competing Pauli symbol
\(Y\). If \(\tau=0.25\), then
\[
\alpha_7=\Delta_7+\tau=0.65.
\]
Since \(e^X(Y)=1\) and \(e^Z(Y)=1\), the perturbed pair is
\[
\bar L_7^X=L_7^X-0.65,
\qquad
\bar L_7^Z=L_7^Z-0.65.
\]
The decoder then recomputes the BP messages incident on \(v_7\) using
\((\bar L_7^X,\bar L_7^Z)\), sets the Pauli symbol at \(v_7\) to \(Y\), updates
the induced binary components, and refreshes the affected residual mismatches
and local RL states. The alternative child receives the metric
\[
M^{(\ell,2)}=M^{(\ell)}+0.40.
\]

Only local quantities need to be refreshed after this expansion. For the present
Tanner graph,
\[
\mathcal N(c_{X,1})=\{v_1,v_3,v_7,v_{10}\},
\qquad
\mathcal N(c_{X,4})=\{v_4,v_6,v_7,v_{10}\},
\]
\[
\mathcal N(c_{Z,1})=\{v_1,v_4,v_7,v_8\},
\qquad
\mathcal N(c_{Z,3})=\{v_3,v_6,v_7,v_9\}.
\]
Since changing \(v_7\) can affect only the incident checks and the variable nodes
adjacent to those checks, the affected second-order neighborhood is
\[
\begin{aligned}
\mathcal V_{\mathrm{aff}}(v_7)
&=
\mathcal N(c_{X,1})\cup\mathcal N(c_{X,4})
\cup\mathcal N(c_{Z,1})\cup\mathcal N(c_{Z,3})\\
&=\{v_1,v_3,v_4,v_6,v_7,v_8,v_9,v_{10}\}.
\end{aligned}
\]
All other variable nodes keep the same local RL state in this expansion step.

Finally, suppose that after the update the ordinary child has residual weight
\(w^{(\ell,1)}=2\), while the alternative child has residual weight
\(w^{(\ell,2)}=1\). Both children are inserted into the temporary candidate set
\(\mathcal C\). If the list size is large enough, both can survive and be
continued. If pruning is required, the decoder compares candidates using the
lexicographic pair \((M,w)\): the ordinary child has smaller metric, while the
alternative child has smaller mismatch weight. Thus, the metric controls how
much the decoder is allowed to deviate from the locally most likely trajectory,
and the mismatch weight is used to break ties among paths with the same metric.

This example highlights the main difference from RL-S. The RL-S decoder would
keep only the ordinary continuation at \(v_7\) and would proceed with a single
trajectory. In contrast, RL-LS keeps the ordinary RL-S continuation while also
allowing a second, softly perturbed continuation toward the nearest competing
Pauli symbol. Therefore, the proposed decoder preserves the learned RL-S path
but adds controlled local path diversity inside the sequential update process.
\end{example}

\section{Complexity and Latency}
\label{sec:Implementation}

We measure complexity and latency in terms of the number of local variable-node updates. In conventional flooding BP, all node updates within one iteration can in principle be executed in parallel, so if the maximum number of iterations is \(I_{\max}\), its decoding latency scales as \(O(I_{\max})\) under ideal parallel hardware. For sparse Tanner graphs with bounded node degrees, its computational complexity also scales as \(O(I_{\max}n)\).

In contrast, both RL-S and RL-LS apply variable-node updates sequentially within each iteration. If one local variable-node update is counted as one unit of work, then the RL-S decoder visits \(n\) variable nodes per iteration, and its worst-case computational complexity is \(O(I_{\max}n)\). Under the same serial-update assumption, its latency is of the same order, namely \(O(I_{\max}n)\).

For our proposed RL-LS decoder, the list size is capped at \(L\), and for each surviving trajectory and selected variable node the decoder forms at most two local continuations: the ordinary RL-S continuation and one alternative continuation. Therefore, in the worst case, one iteration involves at most \(2Ln\) local updates, which gives the overall complexity bound
\[
O(2LI_{\max}n)=O(LI_{\max}n).
\]
When the \(L\) list elements are processed in parallel, then the latency is reduced toward \(O(I_{\max}n)\), but it still remains larger than that of flooding BP because the variable-node updates within each trajectory are inherently sequential. The factor \(2\) above is conservative, since the alternative continuation reuses post-update local quantities already computed for the ordinary continuation, and the cost of ranking and pruning at most \(2L\) candidates per expansion step is a lower-order overhead for the list sizes considered in practice.

Finally, these are worst-case bounds. In practice, as shown in our numerical results, the sequential decoders often require a substantially smaller average number of iterations to converge, and even under the same cap \(I_{\max}\), they typically achieve significantly better error-correction performance than conventional flooding BP.
Moreover, cluster-based variants of sequential scheduling can update multiple variable nodes as a group, thereby reducing the effective sequential depth with negligible degradation in error-correction performance. Such clustered implementations provide a practical path toward bringing the latency of sequential decoders closer to that of conventional BP decoders while retaining the benefits of learned scheduling.

\section{Numerical Results}
\label{sec:Numerical}

In our experiments, the reinforcement-learning used by our proposed RL-LS decoder follows the RL-S framework in~\cite{moradi2026rlbp}. In particular, we use the same learned scheduling policy and offline training procedure as in the RL-S decoder, and the learned Q-table is fixed during inference. 
The difference here lies in the decoding stage: instead of following only a single learned sequential trajectory, the proposed RL-LS decoder preserves the ordinary RL-S continuation and, at each selected variable-node update, can additionally explore a competing branch obtained by softly biasing the post-update LLR pair toward the second-most likely Pauli symbol, recomputing the incident local BP messages, and committing the visited variable node to that second-best symbol.
The resulting candidate trajectories are ranked using our proposed path metric function introduced in Section~III, and only the best \(L\) candidates are retained throughout decoding. The performance is reported in terms of block error rate over the depolarizing channel.

\begin{figure}[t]
  \centering
  \includegraphics[width=\linewidth]{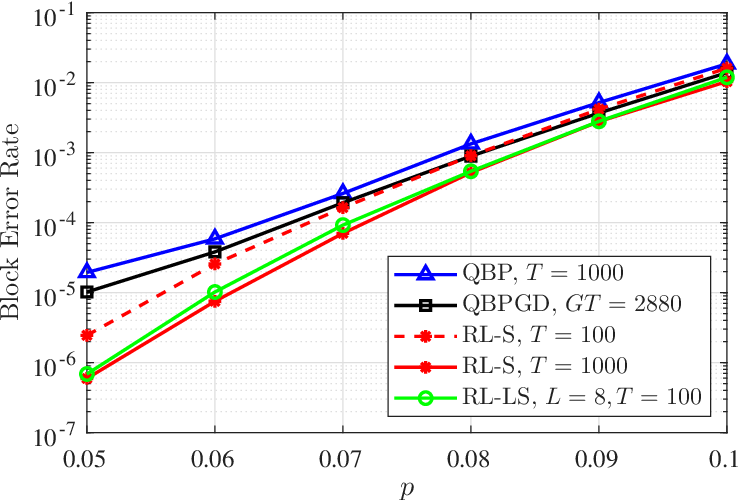}
  \caption{Comparison of the block error rate performance of our proposed RL-LS decoder with other decoders for the \( [[288,12,18]] \) BB code over the depolarizing channel.}
  \label{fig:BB288_RL_LS}
\end{figure}

Fig.~\ref{fig:BB288_RL_LS} shows the error-correction performance of our proposed RL-LS decoder with list size \(L=8\) and maximum number of iterations \(T=100\) for the \( [[288,12,18]] \) BB code introduced in \cite{bravyi2024high}. For comparison, we also include the performance of quaternary BP (QBP), QBP with guided decimation (QBPGD), and RL-S, where RL-S corresponds to the special case of list size \(L=1\), under different iteration settings. For QBPGD, the inner BP decoder uses \(T=10\) maximum iterations, and the maximum number of global iterations (GT) for the block length \(n=288\) is set to \(2880\). The learning-based decoders outperform the conventional BP-based decoders. 
In particular, our proposed RL-LS decoder with \(T=100\) achieves performance comparable to RL-S with \(T=1000\), while requiring only one-tenth of the iteration budget and therefore offering a substantial latency advantage.

\begin{table*}[t]
\centering
\caption{Average number of iterations for the decoding schemes corresponding to Fig.~\ref{fig:BB288_RL_LS} for the \( [[288,12,18]] \) BB code over the depolarizing channel.}
\label{tab:avg_iter_BB288}
\begin{tabular}{|c||c|c|c|c|c|c|}
\hline
$p$ & 0.05 & 0.06 & 0.07 & 0.08 & 0.09 & 0.10 \\ \hline\hline
QBP, $T=1000$             & 3.96 & 5.15 & 6.84 & 9.77 & 16.66 & 31.64 \\ \hline
QBPGD, $\mathrm{GT}=2880$ & 4.08 & 5.46 & 7.53 & 11.71 & 24.94 & 59.84 \\ \hline
RL-S, $T=100$             & 1.94 & 2.21 & 2.55 & 3.09 & 4.07 & 5.95 \\ \hline
RL-S, $T=1000$            & 1.94 & 2.21 & 2.61 & 3.63 & 6.72 & 16.02 \\ \hline
RL-LS, $L=8$, $T=100$     & 1.75 & 2.02 & 2.31 & 2.79 & 3.50 & 5.01 \\ \hline
\end{tabular}
\end{table*}

Table~\ref{tab:avg_iter_BB288} reports the corresponding average number of iterations for the decoding schemes shown in Fig.~\ref{fig:BB288_RL_LS}.
As shown in the Table, the learning-based decoders require substantially fewer average iterations than the conventional BP-based decoders across the whole range of depolarizing probabilities.  
Since the block error rate performance of RL-LS with $L=8$ and $T=100$ is close to that of RL-S with $T=1000$, the results indicate that RL-LS can attain a similar error-correction performance with comparable latency while benefiting from list-based exploration.

\begin{figure}[t]
  \centering
  \includegraphics[width=\linewidth]{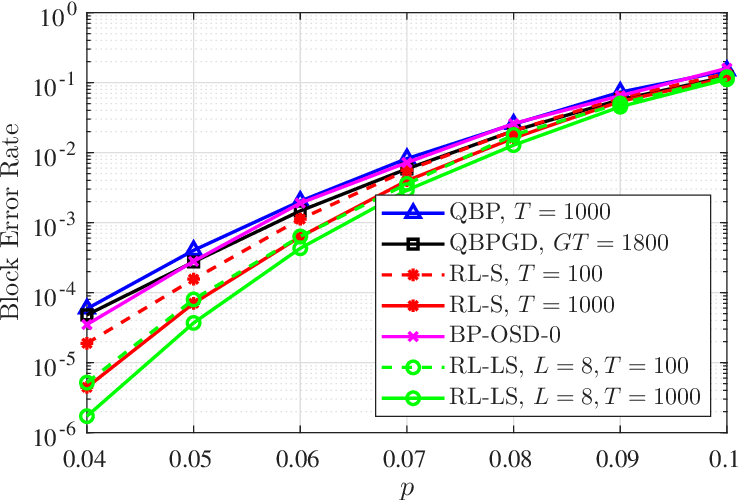}
  \caption{Block error rate comparison of our proposed RL-LS decoder and benchmark decoders for the \( [[180,10,15\leq d \leq 18]] \) A5 code over the depolarizing channel.}
  \label{fig:A5_RL_LS}
\end{figure}

Fig.~\ref{fig:A5_RL_LS} illustrates the block error rate performance of our proposed RL-LS decoder for the \( [[180,10,15\leq d \leq 18]] \) A5 code introduced in \cite{panteleev2021degenerate} over the depolarizing channel. We consider two settings for RL-LS, namely \(L=8\) with \(T=100\) and \(T=1000\), and compare them with QBP, QBPGD, BP-OSD-0, and RL-S under different iteration budgets. The results show that our proposed RL-LS decoder provides the best overall performance among the considered decoders across most of the examined depolarizing crossover probabilities. In particular, the gain is more pronounced in the low- and moderate-noise regimes, where RL-LS achieves a clear reduction in block error rate compared with the other BP-based schemes. Moreover, even with the smaller iteration budget \(T=100\), RL-LS remains competitive with RL-S with \(T=1000\). 
This demonstrates that combining learned sequential scheduling with list-based exploration can substantially improve error-correction performance while maintaining a favorable decoding-latency profile.

\begin{table*}[t]
\centering
\caption{Average number of iterations for the decoding schemes corresponding to Fig.~\ref{fig:A5_RL_LS} for the $\bigl[\!\bigl[180,10,15 \leq d \leq 18\bigr]\!\bigr]$ A5 code over the depolarizing channel.}
\label{tab:avg_iter_A5}
\begin{tabular}{|c||c|c|c|c|c|c|c|}
\hline
$p$ & 0.04 & 0.05 & 0.06 & 0.07 & 0.08 & 0.09 & 0.10 \\ \hline\hline
QBP, $T=1000$             & 2.43 & 3.38 & 6.26 & 14.06 & 36.85 & 79.41 & 166.38 \\ \hline
QBPGD, $\mathrm{GT}=1800$ & 2.44 & 3.57 & 7.69 & 19.09 & 49.90 & 126.04 & 257.66 \\ \hline
RL-S, $T=100$             & 1.61 & 1.88 & 2.37 & 3.34 & 5.54 & 10.27 & 19.11 \\ \hline
RL-S, $T=1000$            & 1.61 & 1.96 & 3.12 & 7.21 & 20.49 & 61.02 & 135.31 \\ \hline
RL-LS, $L=8$, $T=100$     & 1.21 & 1.50 & 1.97 & 2.76 & 4.61 & 8.90 & 15.91 \\ \hline
RL-LS, $L=8$, $T=1000$    & 1.21 & 1.53 & 2.39 & 5.62 & 17.33 & 52.15 & 113.42 \\ \hline
\end{tabular}
\end{table*}

\begin{table*}[t]
\centering
\caption{Average number of iterations for the decoding schemes corresponding to Fig.~\ref{fig:BB144_RL_LS} for the $\bigl[\!\bigl[144,12,12\bigr]\!\bigr]$ BB code over the depolarizing channel.}
\label{tab:avg_iter_BB144}
\begin{tabular}{|c||c|c|c|c|c|c|c|c|}
\hline
$p$ & 0.03 & 0.04 & 0.05 & 0.06 & 0.07 & 0.08 & 0.09 & 0.10 \\ \hline\hline
QBP, $T=100$              & 1.75 & 2.32 & 3.06 & 4.00 & 5.37 & 7.56 & 10.87 & 16.02 \\ \hline
RL-S, $T=10$              & 1.27 & 1.45 & 1.67 & 1.94 & 2.28 & 2.73 & 3.38 & 4.11 \\ \hline
RL-LS, $L=16$, $T=10$     & 1.04 & 1.14 & 1.30 & 1.57 & 1.90 & 2.33 & 2.88 & 3.56 \\ \hline
\end{tabular}
\end{table*}

Table~\ref{tab:avg_iter_A5} reports the corresponding average number of iterations for the decoding schemes shown in Fig.~\ref{fig:A5_RL_LS}.
As shown in Table~\ref{tab:avg_iter_A5}, our proposed RL-LS decoder requires fewer average iterations than the conventional BP-based decoders across the entire range of depolarizing probabilities. In particular, RL-LS with $L=8$ and $T=100$ achieves the smallest average iteration count among the reported schemes, while also providing the strongest block error rate performance in Fig.~\ref{fig:A5_RL_LS}. Even when the maximum iteration budget is increased to $T=1000$, RL-LS still maintains a noticeably smaller average number of iterations than QBP, QBPGD, and RL-S with the same iteration budget, especially in the moderate- and high-noise regimes. These results show that the gain of RL-LS is not only in error-correction performance, but also in decoding convergence speed, since the learned sequential policy together with list-based exploration enables the decoder to converge in fewer iterations on average while achieving better reliability.

\begin{figure}[t]
  \centering
  \includegraphics[width=\linewidth]{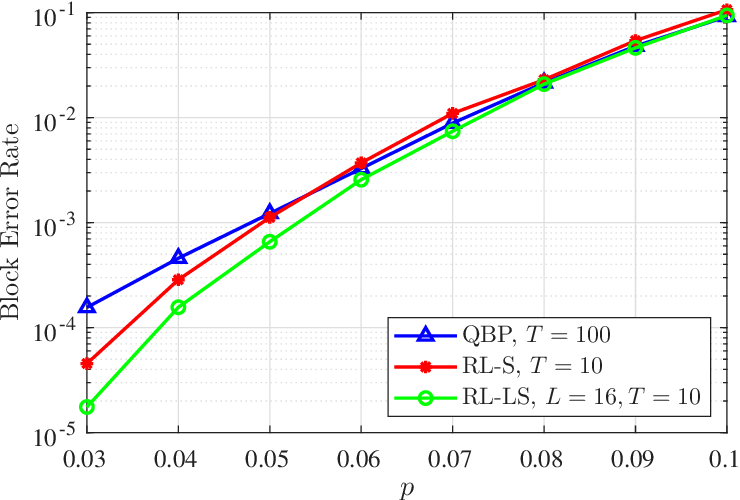}
  \caption{Comparison of the block error rate performance of our proposed RL-LS decoder with other decoders for the \( [[144,12,12]] \) BB code over the depolarizing channel.}
  \label{fig:BB144_RL_LS}
\end{figure}

Fig.~\ref{fig:BB144_RL_LS} presents the block error rate performance of our proposed RL-LS decoder with list size \(L=16\) and maximum number of iterations \(T=10\) for the \( [[144,12,12]] \) BB code introduced in \cite{bravyi2024high}. For comparison, we also show the performance of QBP and RL-S, where RL-S corresponds to the case \(L=1\). It can be observed that the learning-based decoders perform better than the conventional QBP decoder at low depolarizing crossover probabilities. Moreover, under the same iteration budget \(T=10\), our proposed RL-LS decoder achieves more than a twofold reduction in block error rate compared with RL-S at \(p=0.04\), demonstrating the benefit of incorporating list-based search into the learned sequential decoding framework. Since both RL-S and RL-LS operate with the same maximum number of iterations, their decoding latency remains comparable, while RL-LS provides better error-correction performance.

Table~\ref{tab:avg_iter_BB144} reports the corresponding average number of iterations for the decoding schemes shown in Fig.~\ref{fig:BB144_RL_LS}.
As shown in Table~\ref{tab:avg_iter_BB144}, the learning-based decoders require substantially fewer average iterations than the conventional QBP decoder across the entire range of depolarizing probabilities. In particular, our proposed RL-LS decoder with $L=16$ and $T=10$ achieves the smallest average iteration count at all reported channel points, while also providing the best block error rate performance in Fig.~\ref{fig:BB144_RL_LS}. Compared with RL-S under the same iteration budget $T=10$, RL-LS consistently converges in fewer iterations on average, which indicates that the list-based exploration not only improves the error-correction performance but also accelerates decoding convergence. These results further support the advantage of combining learned sequential scheduling with list-based search, since RL-LS attains better reliability without increasing the maximum iteration budget relative to RL-S.

\section{Conclusions}
\label{sec:conclusion}

In this paper, we proposed a learning-based list sequential (RL-LS) belief-propagation decoder for QLDPC codes over the depolarizing channel. Our proposed RL-LS decoder extends the RL-S framework by retaining the ordinary learned sequential decoding path while also exploring a competing path through a soft perturbation toward the second-most likely local Pauli decision. By ranking and pruning candidate trajectories using our proposed cumulative path metric, our decoder is able to better handle local ambiguities caused by short cycles, degeneracy, and unreliable early decisions, while remaining fully within the BP message-passing framework. Numerical results on representative BB and A5 codes showed that our proposed decoder improves upon the underlying RL-S decoder and compares favorably with existing BP-based decoding methods. Overall, the results suggest that combining learned sequential scheduling with list exploration is a promising direction for improving the reliability of BP-based QLDPC decoding. Future work may include adaptive list-size selection, improved path metrics, and combining the proposed approach with other strong BP enhancements for larger codes and other noise models.


\bibliographystyle{IEEEtran}
\bibliography{bibliography}

\end{document}